\documentstyle[preprint,aps,epsfig]{revtex}
\draft
\begin{document}
\preprint{}
\title{Probing the isospin dependence of the in-medium nucleon-nucleon 
cross sections with radioactive beams}
\bigskip
\author{Bao-An Li$^{1,2}$, Pawel Danielewicz$^{1}$, and William G. Lynch$^{1}$ }
\address{$^1$ National Superconducting Cyclotron Laboratory and 
Department of Physics,\\ Michigan State University, East Lansing, MI 48824, USA\\
$^2$ Department of Chemistry and Physics, P.O. Box 419,\\
Arkansas State University, State University, Arkansas 72467-0419, USA}
\maketitle

\begin{quote}
Within a transport model we search for potential probes of the isospin 
dependence of the in-medium nucleon-nucleon (NN) cross sections. 
Traditional measures of the nuclear stopping power are found
sensitive to the magnitude but they are ambiguous for determining the isospin dependence of 
the in-medium NN cross sections. It is shown that isospin tracers, such as 
the neutron/proton ratio of free nucleons, at backward rapidities/angles 
in nuclear reactions induced by radioactive beams in inverse kinematics is 
a sensitive probe of the isospin dependence of the in-medium 
NN cross sections. At forward rapidities/angles, on the other hand, they are
more sensitive to the density dependence of the symmetry energy.   
Measurements of the rapidity/angular dependence of the isospin transport
in nuclear reactions will enable a better understanding of the 
isospin dependence of in-medium nuclear effective interactions. 
\\ 
{\bf PACS} numbers: 25.70.-z, 25.75.Ld., 24.10.Lx
\end{quote}

\newpage
\section{Introduction}
The isospin dependence of in-medium nuclear effective interactions is  
critical in determining both the nature of nucleonic matter and novel structures 
of radioactive nuclei\cite{lpr}. Moreover, it determines the equation of state (EOS), 
especially the nuclear symmetry energy, and transport properties of isospin asymmetric 
nuclear matter. It is thus important for our understanding about many 
interesting questions about not only nuclei, but also neutron stars and supernove\cite{lat01,andrew}.  
Nuclear reactions induced by radioactive beams provide a unique opportunity to explore the
isospin dependence of in-medium nuclear effective interactions. 

While much attention has been given to finding experimental observables constraining 
the EOS of isospin asymmetric nuclear matter, little effort has been made so far 
to extract the isospin dependence of the in-medium nucleon-nucleon (NN) cross sections. 
The latter affects the transport properties 
of isospin asymmetric nuclear matter\cite{chen01} and it depends on particularly 
the short-range part of nuclear 
effective interactions. Because both the iso-singlet and iso-triplet channels contribute to 
neutron-proton (np) scatterings, their cross sections ($\sigma_{np}^{free}$) in free space
are higher than those for proton-proton (pp) or neutron-neutron (nn) 
scatterings ($\sigma_{pp}^{free}$) where only iso-triplet channels are involved.
This is illustrated by the solid line in Fig.\ 1 where the ratio $\sigma_{np}/\sigma_{pp}$
is shown as a function of nucleon beam energy $E_{lab}$. More specifically, the 
$\sigma_{np}/\sigma_{pp}$ ratio changes from 
about 2.7 at $E_{lab}=50$ MeV to 1.7 at $E_{lab}=300$ MeV. How does the ratio 
$\sigma_{np}/\sigma_{pp}$ change with density and isospin asymmetry 
in asymmetric medium encounterd often in heavy-ion reactions and astrophysical situations? 
This is an important question since its answer may reveal directly useful information 
about the isospin dependence of the in-medium nuclear effective interactions. However, 
very little work has been done so far about the isospin dependence 
of the in-medium NN cross sections in asymmetric nuclear matter 
although extensive studies have been carried out in symmetric matter based on 
various many-body theories and/or 
phenomenological approaches, see, e.g., refs.\cite{pan,gqli,schu,gale}. Therefore, one can find 
in the literature only some information about the density dependence of the 
$\sigma_{np}/\sigma_{pp}$ ratio in symmetric nuclear matter.
\begin{figure} 
\centering \epsfig{file=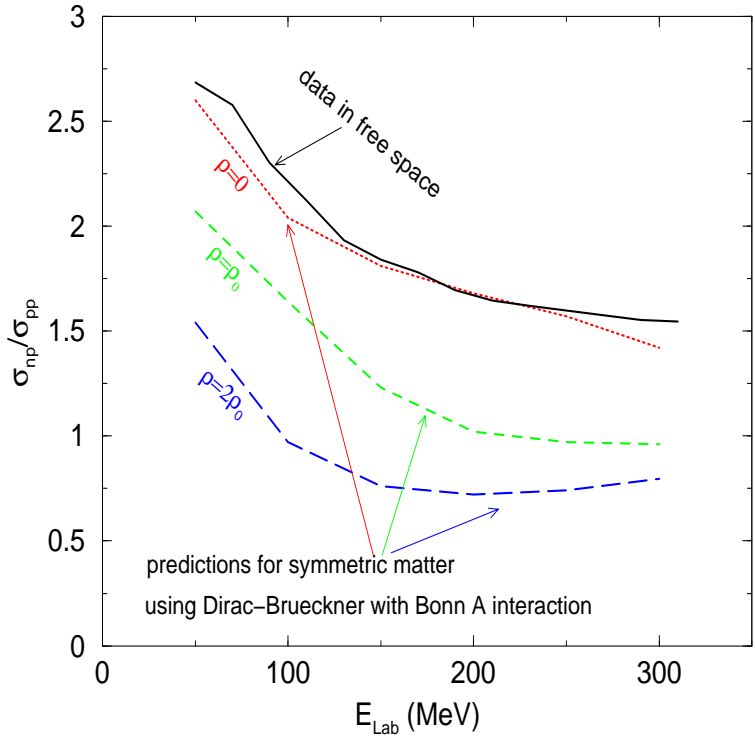,width=10cm,height=10cm,angle=-90}
\vspace{1 cm} 
\caption{(Color online) The ratio of np over pp scattering cross sections 
as a function of incident nucleon energy.
The solid line is extracted from experimental data\protect\cite{data} while the dashed lines are
extracted from calculations in symmetric matter using the Bonn A potential within the 
Dirac-Bruckner approach in ref.\protect\cite{gqli}.} 
\label{fig1} 
\end{figure}       
As an example, shown in Fig. 1 with the dashed lines are the $\sigma_{np}/\sigma_{pp}$ ratio in 
symmetric matter extracted from predictions using the Bonn A potential within the 
Dirac-Brueckner approach of ref.\cite{gqli}. Efforts are currently being made to extend the
above calculations to isospin asymmetric matter within the same approach, and the preliminary results 
are indeed very interesting\cite{idaho}. 
In this approach not only the in-medium NN cross sections are reduced compared to 
their values in free-space, the ratio $\sigma_{np}/\sigma_{pp}$ 
is also predicted to decreases with increasing density. However, several other microscopic studies
have concluded just the opposite, i.e., the $\sigma_{np}/\sigma_{pp}$ ratio increases in symmetric 
medium, see, e.g., \cite{gg,mk,qli}. It is therefore
imperative to have experimental information about the isospin dependence of the in-medium NN cross sections. 
Experimentally, strong evidences supporting reduced in-medium NN cross sections have been
found in heavy-ion collisions, see, e.g., refs.\cite{gar,xu,pawel}. However, 
all analyses have been done so far assume some overall reduction of all NN scattering 
channels. Thus, no information about the isospin dependence of 
the in-medium NN cross sections has been extracted from the experiments. 
Given the opportunities provided by the radioactive beams, it now becomes more important to find 
sensitive experimental observables practically useful for extracting the isospin dependence 
of the in-medium NN cross sections. In this work we demonstrate within a transport 
model that isospin tracers, such as the neutron/proton ratio of free nucleons, at backward 
rapidities/angles in nuclear reactions induced by radioactive beams in inverse kinematics 
is such an observable. 

\section{The transport model and its most important inputs used in this work}
In this exploratory study, we use an {\it isospin dependent but momentum independent}
transport model\cite{ibuu}. This is perfectly sufficient for the purpose of this work 
while being computationally efficient.  The default values of the 
differential and total NN cross sections are taken from the experimental data giving the solid 
line in Fig.\ 1\cite{data,ireview}. We explore effects of the isospin dependence of the 
in-medium NN cross sections by changing the ratio $\sigma_{np}/\sigma_{pp}$ without changing 
the angular distributions of elementary NN scatterings. Besides the NN cross sections, another
input to the model important for the following discussions is the symmetry energy $E_{sym}(\rho)$.   
The density dependence of the symmetry energy is rather strongly model 
dependent, see, e.g., refs.\cite{ireview,ibook,science,stone}. We adopt here a 
parameterization used by Heiselberg 
and Hjorth-Jensen in their studies of neutron stars\cite{hei00}
$E_{sym}(\rho)=E_{sym}(\rho_0)\cdot (\rho/\rho_0)^{\gamma}$,
where $E_{sym}(\rho_0)$ is the symmetry energy 
at normal nuclear matter density $\rho_0$ and $\gamma$ is a parameter. 
By fitting earlier predictions of the variational many-body calculations 
by Akmal et al\cite{akm97}, they obtained the values of 
$E_{sym}(\rho_0)=$32 MeV and $\gamma=0.6$. However, recent analyses of 
isospin diffusions in heavy-ion collisions at intermediate energies 
favor strongly a $\gamma$ value between 1 and 2\cite{betty,chen} depending on whether the 
momentum dependence of the symmetry potential is taken into account. In the
following we use $E_{sym}(\rho_0)=30$ MeV and compare results obtained with 
$\gamma=$ 1 and 2. By construction, the symmetry energies with $\gamma=1$ and $2$ come
to cross each other at $\rho_0$. At subnormal densities the softer symmetry energy with $\gamma=1$
leads to more repulsive/attractive symmetry potentials for neutrons/protons than the
stiffer one with $\gamma=2$. While it is the opposite at supranormal densities. 
The initial nucleon density distributions in the 
projectile $^{100}Zn$ were calculated by using the Hartree-Fock-Bogoliubov
method and were provided to us by J. Dobaczewski\cite{jaeck}. Other details about the model 
can be found in earlier publications\cite{ibuu,ireview}.

\begin{figure}[htp] 
\centering \epsfig{file=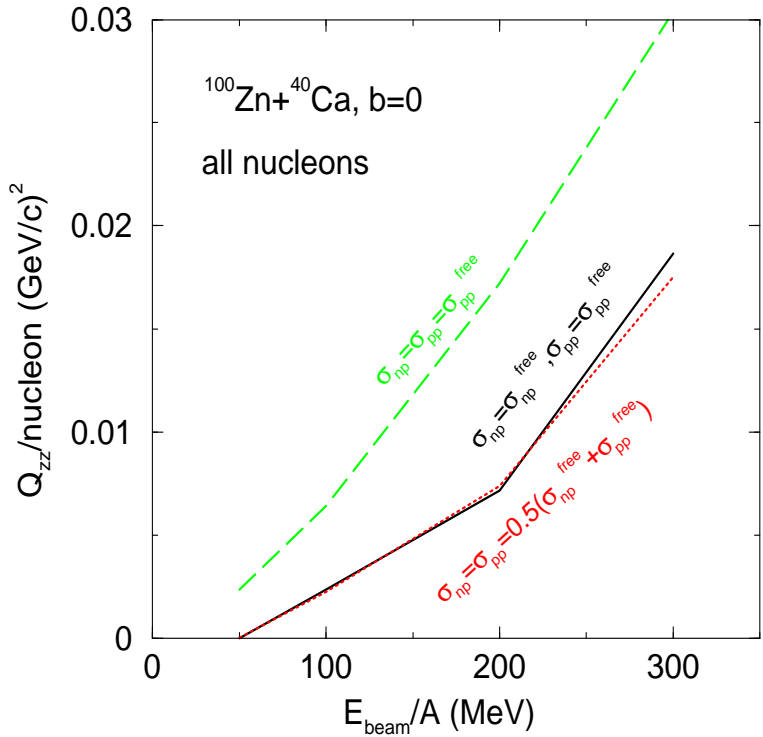,width=10cm,height=10cm,angle=-90} 
\vspace{1.0 cm} 
\caption{(Color online) Quadruple moment as a function of beam energy in head-on collisions
of $^{100}Zn+^{40}Ca$ with the three choices of nucleon-nucleon cross sections.} 
\label{fig2} 
\end{figure}     
\section{The global stopping power as a measure of the isospin dependence of the 
in-medium NN cross sections}
First, it is worth mentioning that we have examined several observables 
that are known to be sensitive to the in-medium NN cross sections. 
These include the quadruple moment $Q_{zz}$ of nucleon momentum distribution,
the linear momentum transfer (LMT) and the ratio of transverse to 
longitudinal energies (ERAT) which have all been used traditionally as 
measures of the nuclear stopping power. We found that these observables are 
sensitive only to the magnitude but not the isospin dependence (measured by 
the $\sigma_{np}/\sigma_{pp}$ ratio) of the in-medium NN cross sections. 
The quadruple moment $Q_{zz}$ was previously proposed by Liu et al. 
as a measure of the isospin dependence of the in-medium NN cross sections based on 
their IQMD model calculations\cite{liu}. This seems to be in contradiction to our findings here.
We thus examine here this measure in detail and discuss the origin of the seemingly different 
conclusions. 

Shown in Fig.\ 2  are the quadruple moment per nucleon
$Q_{zz}/A\equiv \frac{1}{A}\sum^A_{i=1}(2p_{iz}^2-p_{ix}^2-p_{iy}^2)$
as a function of beam energy for the head-on collisions of $^{100}Zn+^{40}Ca$ with 
three choices of the in-medium NN cross sections. In agreement with ref.\cite{liu} 
we found that the $Q_{zz}$ is almost independent of the symmetry energy simply because 
the isoscalar interaction overwhelmingly dominates over the isovector interaction for 
the globe thermalization of the system. Also in agreement with ref.\cite{liu}, by setting
artificially the cross section for neutron-proton scatterings to be the same as that for
proton-proton scatterings in free-space (long dashed line), thus the 
ratio $\sigma_{np}/\sigma_{pp}$ is one, the $Q_{zz}$ increases significantly compared to the 
calculations using the free-space np and pp scattering cross sections $\sigma_{np}^{free}$ and 
$\sigma_{pp}^{free}$ (solid line). Based on this observation, it was proposed in 
ref.\cite{liu} that the stopping power measured by the  $Q_{zz}$ 
can be used as a sensitive probe of the isospin dependence of the in-medium NN 
cross sections. However, we point out that the observed increase of the $Q_{zz}$ 
is simply due to the reduction of the np scattering cross sections 
although the $\sigma_{np}/\sigma_{pp}$ ratio is indeed also changed.  
In fact, the $Q_{zz}$ is insensitive to the $\sigma_{np}/\sigma_{pp}$ 
ratio if one keeps the total number of NN collisions to be about the same. 
We demonstrate the ambiguity of using the $Q_{zz}$ probe by comparing 
the above calculations with the ones using 
$\sigma_{np}=\sigma_{pp}=(\sigma_{np}^{free}+\sigma_{pp}^{free})/2$. In the latter the
ratio $\sigma_{np}/\sigma_{pp}$ is also one, however, the $Q_{zz}$ is about the same as
the calculations using the free-space NN cross sections up to about $E_{beam}/A=220$ MeV.
This observation can be understood qualitatively from the total number of 
NN collisions $N_{coll}$ that essentially determines the nuclear stopping power. 
Neglecting the Pauli blocking, the $N_{coll}$ scales according to 
$N_{coll}\propto N_{np}\sigma_{np}+(N_{pp}+N_{nn})\sigma_{pp}$,
where the $N_{np}$ and $N_{pp}$ are the number of np and pp colliding pairs.
Assuming only the first chance NN collisions contribute, the ratio $N_{np}/(N_{pp}+N_{pp})
\approx (1-\delta_1\delta_2)/(1+\delta_1\delta_2)\approx 1-2\delta_1\delta_2$ 
is about one to the second order in isospin asymmetry even for the very neutron-rich systems, 
where $\delta_1\equiv (N_1-Z_1)/A_1$ and $\delta_2\equiv (N_2-Z_2)/A_2$ are the isospin asymmetries of 
the two colliding nuclei. Thus one has $N_{coll}\propto N_{np}(\sigma_{np}+\sigma_{pp})$.
With either $\sigma_{np}=\sigma_{pp}=(\sigma_{np}^{free}+\sigma_{pp}^{free})/2$ or 
$\sigma_{np}=\sigma_{np}^{free}$ and $\sigma_{pp}=\sigma_{pp}^{free}$ the numbers of 
NN collisions $N_{coll}$ are then the same leading to approximately the same $Q_{zz}$. 
At higher energies, however, secondary collisions are expected to become gradually more important. 
The above arguments become less valid. 
Our discussions here indicates clearly that the nuclear stopping power is indeed 
sensitive to the in-medium NN cross sections. However, the stopping power alone is insufficient
to determine simultaneously both the magnitude and the isospin dependence of the in-medium NN cross sections.
In a nutshell, one needs at least two observables to determine two unknowns. An additional observable 
sensitive to the ratio $\sigma_{np}/\sigma_{pp}$ is thus absolutely necessary.  

\section{The neutron/proton ratio as a measure of the isospin dependence of the 
in-medium NN cross sections}

Now we turn to the rapidity and angular distributions of isospin 
tracers as potential probes of the isospin dependence of the in-medium NN cross sections.
Several observables can be used as isospin tracers, such as, the neutron/proton ratio or 
isospin asymmetry $\delta$ of free nucleons and fragments. The rapidity and angular distributions of the 
isospin tracers measure directly the isospin transport in reactions especially 
below the pion production threshold. 
These observables were previously used also to study the momentum stopping 
power and the nucleon translucency\cite{sherry,bass,lisherry1,liko98,mosel,rami} 
in heavy-ion collisions, see, e.g., \cite{ireview,lisherry2} for a review. 
We use the isospin tracers at backward rapidities/angles in central collisions induced 
by highly asymmetric projectiles on symmetric targets in inverse kinematics to probe 
the isospin dependence of the in-medium NN cross sections. In these reactions 
the deviation of neutron/proton ratio from one at backward rapidities/angles 
reflects the strength of isospin transfer from the projectile to the target. Our proposal 
is based on the consideration that only large angle and/or multiple np scatterings are effective
in transporting the isospin asymmetry from forward to backward angles. 
With inverse kinematics nucleons in the lighter target moving backward with higher velocities
in the center of mass frame of the reaction are more likely to induce 
multiple np scatterings. It is well known that the symmetry potential is also important 
for isospin transport in heavy-ion collisions\cite{far,li97,shi,ditoro}. However, it is unlikely 
for the symmetry potential to change the directions of motion of nucleons. Thus at backward
rapidities/angles, the isospin tracers are less affected by the symmetry potential.  
Nevertheless, the relative importance and interplay of the symmetry 
potential and the in-medium NN cross sections on the rapidity/angular distributions of 
isospin tracers have to be studied quantitatively within a transport approach.
We look for observables in special kinematic or geometrical regions where the dual sensitivity
to both the symmetry potential and the isospin dependence of the in-medium NN cross sections is a
minimum if it can not be avoided completely. 

Shown in Fig. 3 and Fig. 4 are the rapidity distributions of all nucleons (lower windows) 
and their isospin asymmetries (upper windows) at 100 fm/c 
in head-on collisions of $^{100}Zn+^{40}Ca$ at a beam energy of 
200 MeV/A using $\gamma=1$ and $2$, respectively. 
\begin{figure}[htp] 
\centering \epsfig{file=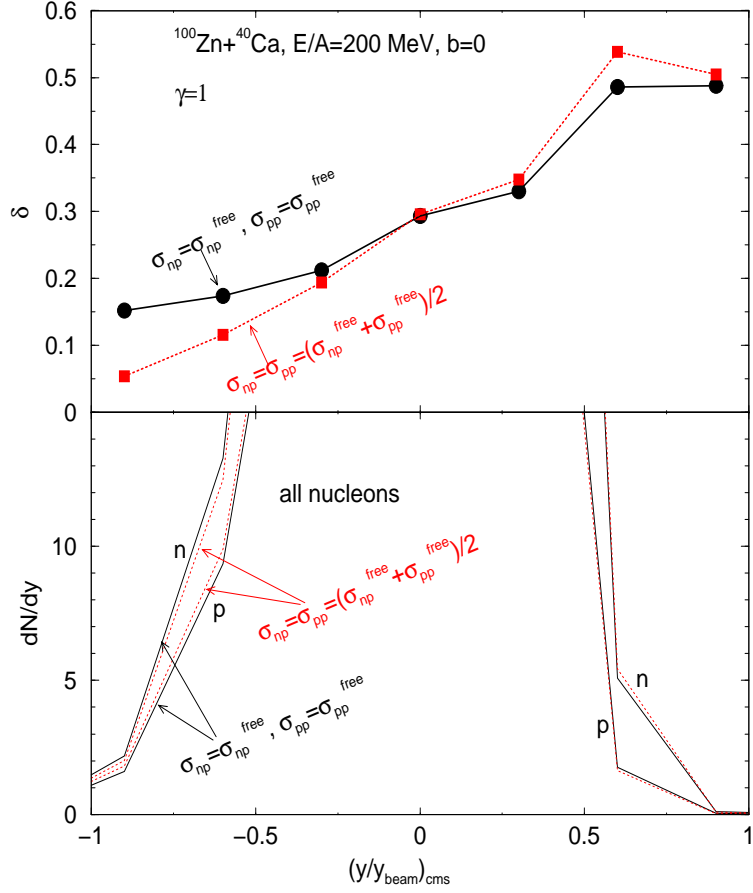,width=12cm,height=10cm,angle=-90} 
\vspace{1.0 cm} 
\caption{(Color online) Rapidity distributions (lower window) of all nucleons and their 
isospin asymmetries (upper window) in head-on collisions of $^{100}Zn+^{40}Ca$ 
at a beam energy of 200 MeV/A using a $\gamma$ parameter of 1.} 
\label{fig3} 
\end{figure}     
We first compare nucleon rapidity distributions 
using the free-space NN cross sections and 
$\sigma_{np}=\sigma_{pp}=(\sigma_{np}^{free}+\sigma_{pp}^{free})/2$. As discussed earlier 
and shown in Fig.\ 2 these two choices of the in-medium NN cross sections lead to identical 
quadruple moment $Q_{zz}$
at $E_{beam}=200$ MeV/A. It is seen that the effects of the in-medium NN cross 
sections on the overall nucleon rapidity distributions 
are rather small with both values of the $\gamma$ parameter. Moreover, the symmetry 
energy also has very little effect on the nucleon rapidity distributions. 
These observations are consistent with those obtained from studying other global 
measures of the nuclear stopping power. Concentrating on the forward and backward nucleons, however, it is
clearly seen that the larger $\sigma_{np}/\sigma_{pp}$ ratio in the case of using
$\sigma_{np}=\sigma_{np}^{free}$ and $\sigma_{pp}=\sigma_{pp}^{free}$ leads to more (less) transfer of
neutrons (protons) from forward to backward rapidities. Since the effect is opposite 
on neutrons and protons, it is much more pronounced on the isospin asymmetry $\delta$ 
as shown in the upper windows. 
\begin{figure}[htp] 
\centering \epsfig{file=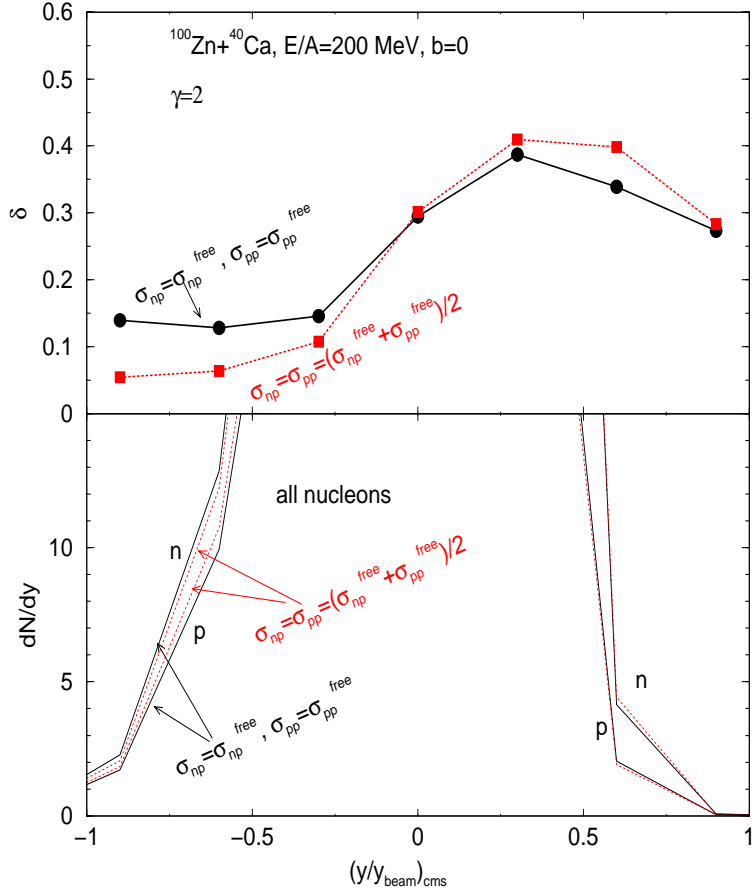,width=12cm,height=10cm,angle=-90} 
\vspace{1.0 cm} 
\caption{(Color online) The same as fig. 3 but using a $\gamma$ parameter of 2.} 
\label{fig4} 
\end{figure}     
It is seen that the isospin asymmetries are rather 
sensitive to the isospin dependence of the in-medium 
NN cross sections especially at backward rapidities in both cases. Comparing the two upper windows 
of Figs. 3 and 4, one can notice a small effect of the symmetry potential especially at forward rapidities.  
At backward rapidities, however, the influence of the isospin dependence of the in-medium NN 
cross sections dominates overwhelmingly over that due to the symmetry potential. 
The effects on $\delta$ due to the isospin dependence of the in-medium NN cross sections 
discussed above are clearly measurable, especially at backward rapidities.
In principle, the effect can be extracted experimentally, for instance, by studying the 
free neutron/proton ratio, the $t/^3He$ ratio or the isoscaling parameters. 
As an illustration, we now turn to the polar angle distributions of  
the neutron/proton ratio $(n/p)_{free}$ of free nucleons identified as 
those having local baryon densities less than $\rho_0/8$. 
\begin{figure}[htp] 
\centering \epsfig{file=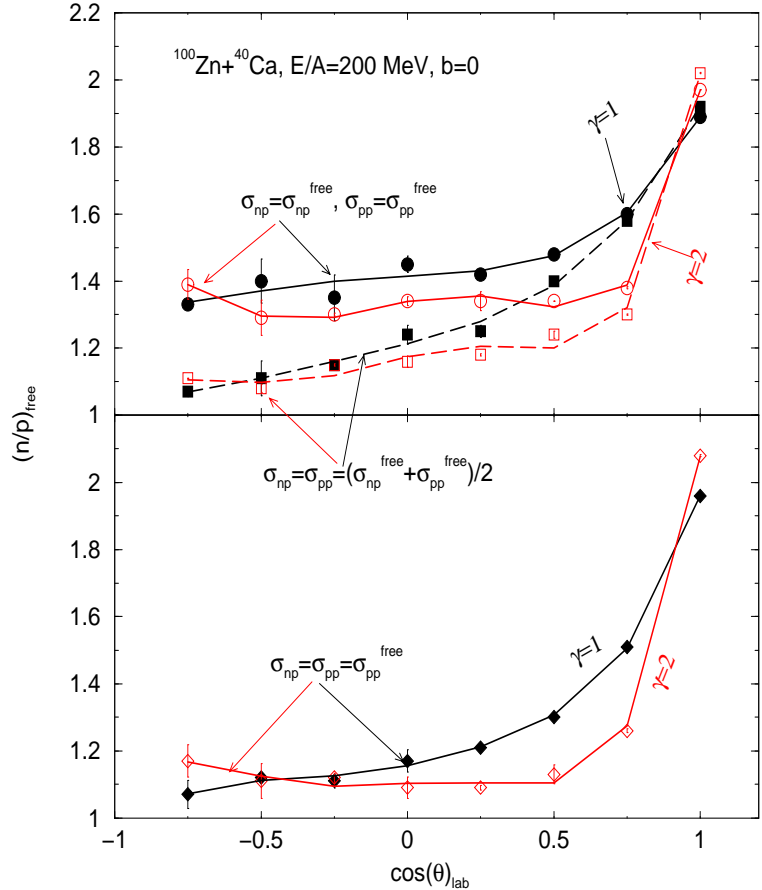,width=12cm,height=10cm,angle=-90} 
\vspace{1.0 cm} 
\caption{(Color online) Angular distributions of the free neutron to proton ratio $(n/p)_{free}$ 
in head-on collisions of $^{100}Zn+^{40}Ca$ at a beam energy of 200 MeV/A.} 
\label{fig5} 
\end{figure}     
These are shown in 
Fig. 5 for the three choices of the in-medium NN cross sections with both $\gamma=1$ and $2$. 
In the upper window we compare results obtained by using the free-space NN cross sections 
and the choice $\sigma_{np}=\sigma_{pp}=(\sigma_{np}^{free}+\sigma_{pp}^{free})/2$, 
the same choices as those in Figs.\ 3 and 4. 
It is clearly seen that the $(n/p)_{free}$ ratio at backward angles is rather 
insensitive to the symmetry energy but very sensitive to the isospin dependence 
of the in-medium NN cross sections. While at forward angles it is the opposite. 
Moreover, by comparing results using all three choices considered for 
the in-medium NN cross sections, it is interesting to see that the choices of 
$\sigma_{np}=\sigma_{pp}=(\sigma_{np}^{free}+\sigma_{pp}^{free})/2$ and 
$\sigma_{np}=\sigma_{pp}=\sigma_{pp}^{free}$ lead to about the 
same $(n/p)_{free}$ value at very backward angles. The latter value 
is significantly less than the one obtained by using the free np and pp cross sections.  
In other words, at these very backward angles the $(n/p)_{free}$ is sensitive only to the    
$\sigma_{np}/\sigma_{pp}$ ratio but not the absolute values of the individual
nn and np cross sections nor the symmetry energies. Thus it would be very valuable 
to measure the $(n/p)_{free}$ ratio at large backward angles. On the other hand, 
at very forward angles the $(n/p)_{free}$ ratio is very sensitive to the symmetry
potential but not much to the in-medium NN cross sections.        

To further illustrate and test our proposal we study in Fig.\ 6 the  
$(n/p)_{free}$ ratio as a function of nucleon kinetic energy in the laboratory frame.
We set a limit of $cos(\theta)\leq -0.25$ for backward (upper window) 
and $cos(\theta)> 0.5$ for forward (lower window) angles. Most nucleons emitted to
the backward angles have energies less than about 100 MeV for the reaction considered. 
Only few nucleons in the backward regions have higher energies and our calculations
using 12,000 events in each case do not have enough statistics to show a 
meaningful $(n/p)_{free}$ ratio. In the backward angles the $(n/p)_{free}$ ratio is
significantly higher than one which is the neutron/proton ratio of the 
target considered here. The value of $(n/p)_{free}$ is larger with the higher 
$\sigma_{np}/\sigma_{pp}$ ratio and the effect of the isospin dependence of the in-medium NN 
cross section is most pronounced at very low energies. This is understandable 
because transferring relatively more 
neutrons from the forward-going projectile to the backward direction 
requires more np scatterings.
\begin{figure}[htp] 
\centering \epsfig{file=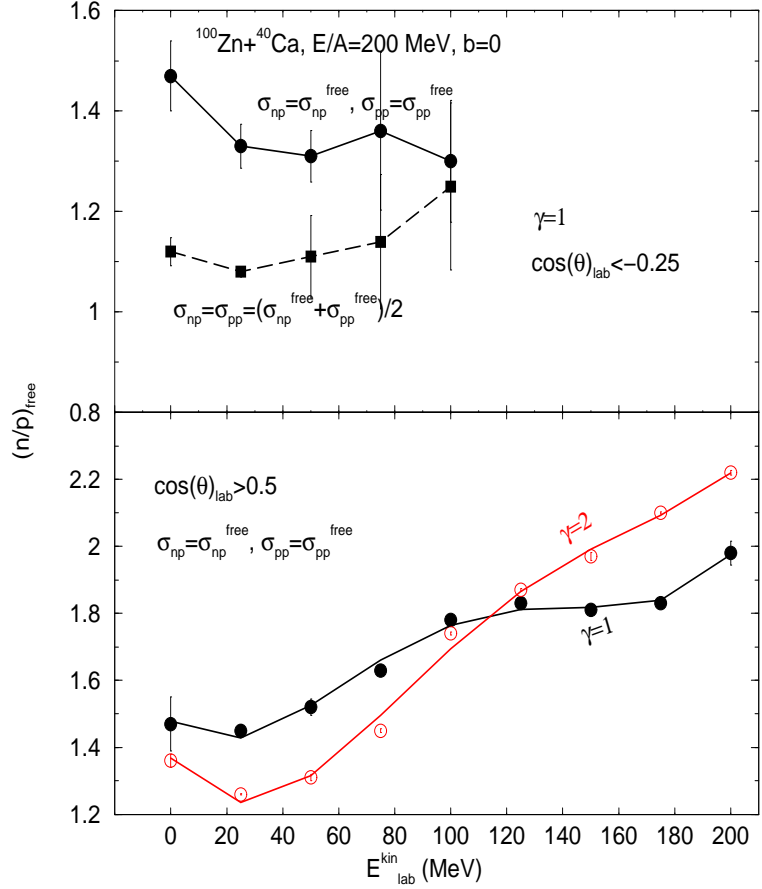,width=12cm,height=10cm,angle=-90}
\vspace{1.0 cm}  
\caption{(Color online) The $(n/p)_{free}$ as a function of nucleon kinetic energy at
backward (upper window) and forward (lower window) angles in head-on collisions of 
$^{100}Zn+^{40}Ca$ at a beam energy of 200 MeV/A.} 
\label{fig6} 
\end{figure}     
Once these neutrons are converted 
backward through possibly multiple scatterings they then have less energies left. 
Moreover, since these neutrons have experiences multiple np scatterings they are therefore 
more sensitive to the $\sigma_{np}/\sigma_{pp}$ ratio. In the forward angles selected here
the $(n/p)_{free}$ ratio is more affected by the symmetry energy. As an example, we show in the 
lower window of Fig.\ 6 the results obtained by using the free NN cross sections. 
Results with other choices of the in-medium
NN cross sections are qualitatively the same. But with the still relatively large 
angular range of $-60^0\leq \theta \leq 60^0$ selected by the cut $cos(\theta)>0.5$, 
the in-medium NN cross sections still
have some effects on the $(n/p)_{free}$ ratio at forward angles 
as indicated in Fig.\ 5. It is seen that 
the influence of the symmetry energy depends strongly on the nucleon energy as one expects. 
Since the low energy nucleons are more likely emitted at subnormal densities where the 
repulsive/attractive symmetry potentials are stronger with the softer symmetry energy, 
the $(n/p)_{free}$ ratio is higher with $\gamma=1$ for low energy nucleons. 
The high energy nucleons mostly emitted forward, however, more likely have gone through 
the supranormal density region in the earlier stage of the reaction. The 
stiffer symmetry energy with $\gamma=2$ thus results in higher 
values of $(n/p)_{free}$ for these nucleons.     

\begin{figure}[htp] 
\centering \epsfig{file=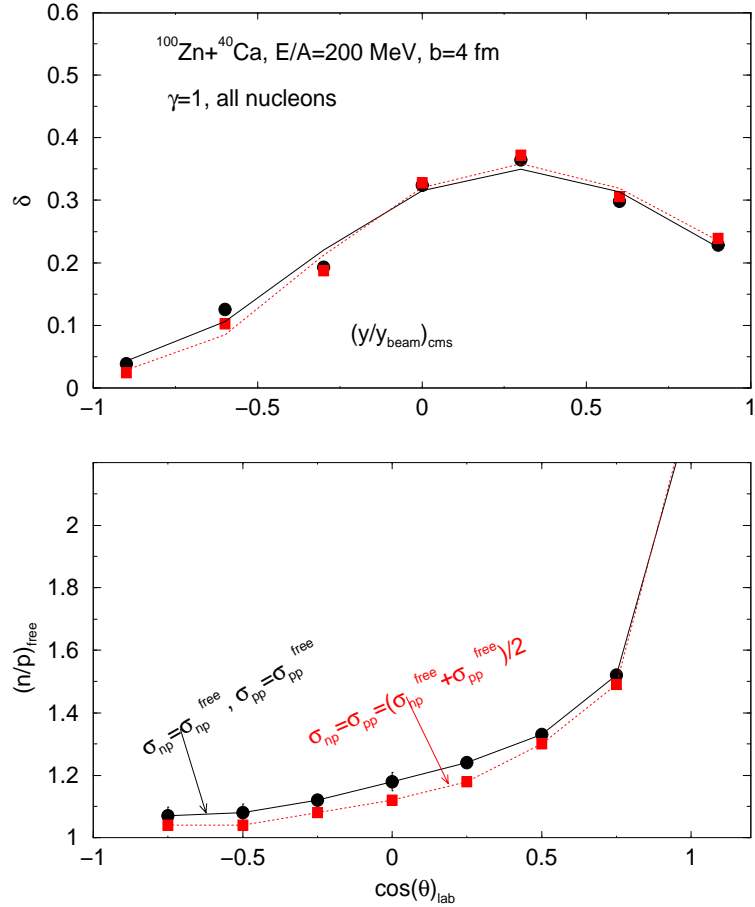,width=12cm,height=10cm,angle=-90}
\vspace{1.0 cm}  
\caption{(Color online) Rapidity (upper window) and angular (lower window) 
dependences of isospin asymmetries of all nucleons in the reaction of $^{100}Zn+^{40}Ca$ 
at a beam energy of 200 MeV/A and impact parameter of 4 fm using a $\gamma$ parameter of 1.} 
\label{fig7} 
\end{figure}       
The above discussions are all based on results of head-on collisions. We find that 
the conclusions remain qualitatively the same but with reduced effects at finite impact
parameters. As an example, shown in Fig. 7 are the rapidity and angular distributions of the  
isospin asymmetry $\delta$ of all nucleons (upper window) and the $(n/p)_{free}$ of free ones 
at an impact parameter of 4 fm. The effects of the in-medium NN cross sections 
are still clearly observable but smaller than those in head-ion collisions. One 
can also notice that memories of the n/p ratio of the projectile and target are now more clear 
as one expects. Besides the reactions at 200 MeV/A, we have also studied the 
reactions at 100 and 300 MeV/A but with less events so far. Our conclusions do not 
seem to change qualitatively in the energy range considered. It will be interesting 
to extend the study down to the Fermi energy range and examine the influence of the size of the 
colliding nuclei. Also, an investigation based on a {\it momentum dependent} 
transport model\cite{li04} using isospin-dependent in-medium NN cross sections and the mean field 
evaluated consistently in asymmetric matter from the same nuclear effective 
interactions\cite{das} is in progress.      

\section{Summary}
In summary, within a transport model for nuclear reactions 
induced by neutron-rich nuclei, we searched for potential probes of 
the isospin dependence of the in-medium NN cross sections. The traditional probes of the 
nuclear stopping power are found sensitive to the magnitude of the in-medium NN cross sections. 
They are, however, ambiguous for determining the isospin dependence of the in-medium NN cross sections.  
In particular, we found that an earlier conclusion that the nucleon quadruple moment can be 
used as a probe of the isospin dependence of the in-medium NN cross sections is premature.
We also studied the relative importance and interplay of the symmetry energy and the 
in-medium NN cross sections on the rapidity and angular distributions of isospin tracers.
We found that the isospin tracers, such as the neutron/proton ratio of free nucleons, 
at backward rapidities/angles in nuclear reactions induced by radioactive beams in 
inverse kinematics is a sensitive probe of the isospin dependence of the in-medium NN cross sections. 
At forward rapidities/angles, on the other hand, the neutron/proton ratio is more 
sensitive to the density dependence of the symmetry energy. It is thus very useful to 
measure experimentally the rapidity and angular distributions of isospin tracers to study 
the transport properties and the EOS of isospin asymmetric matter. Ultimately,
these studies will enable us to better understand the isospin dependence of the in-medium
nuclear effective interactions.   

B.A. Li would like to thank the nuclear theory group at Michigan State University for 
the kind hospitality he received there during his summer visit when this work started.   
The work is supported in part by the National Science Foundation under Grant No. 
PHY-0245009, PHY-01-10253, PHY-0354572, and the NASA-Arkansas Space Grants Consortium.

\end{document}